# Work Integrated Learning (WIL) In Virtual Reality (VR)

Waleed Al Shehri

Department of Computing, Macquarie University

Sydney, NSW 2109, Australia

*waleed.alshehri@students.mq.edu.au*

**Abstract**

The ultimate goal of industrialized, developed economies is to increase economic growth and to have a highly-skilled, flexible and responsive workforce in order to remain competitive in an increasingly global economic context. Therefore, there is increasing focus on exploring the various options for optimizing worker training. Apprenticeships, traineeships and internships are some examples of the applications of Work Integrated Learning (WIL) with various aims such as to enhance learner skills, provide learners with real life facilities to experience the work environment in which they have to perform in the near future and to enhance their capabilities to handle real time problems, even before they are exposed to them in the workplace. However, technology is increasingly replacing the concept of traditional learning, particularly through innovative IT approaches such as Virtual Reality (VR). Integrating both WIL and VR offers the opportunity to apply WIL in a virtual rather than real work environment as "Virtual Work Integrated Learning (VWIL)". The focus of this report is to initially discuss the concepts WIL and VR, their main characteristics and current applications. Moreover, the pros and cons of VWIL are also analyzed. Finally, the report presents some recommendation including further researches into areas where VWIL has potential to be successful in the future.

***Keywords:*** *Work Integrated Learning, Virtual Reality, Technology, Industry, Education, Workforce**,** Stakeholders.*

.

# 1. Introduction

It has been a long standing desire of modern industrialized societies to streamline education and training programs in order to provide well trained, responsive, highly skilled and experienced workforces. Key to that objective is the potential for integrating academic learning with the practical work environment, in order to improve the learning / training process and increases its efficiency. Work Integrated Learning (WIL) has been introduced, as a modern, formal and structured approach for achieve this objective (Smith et al., 2010). At the core of WIL, the purpose is to follow an integrated two track approach, allowing the leaner to follow an academic curriculum while at the same time including exposure to and practice within an actual work environment. Meanwhile, Virtual Reality (VR), as computer-generated environments simulating real physical environments and physical situations, are been increasingly suggested as highly complementary to WIL (Goold & Augar, 2009).

The concept of Work Integrated Learning has evolved significantly recently and it is now practiced in leading institutions around the world, aiming to provide learning opportunities coupled with real world experiences in one integrated package. Universities, for instance currently tend not only to teach their students pure theoretical knowledge, but also to equip them with relevant skills that can help them achieve success in practical situations (McIlveen et al., 2011).

On the other hand, Virtual Reality is an IT technique used in vastly diverse contexts, ranging from engineering design, to medical applications and even highly sophisticated military operations. It relies on simulating the real environment in order to give impression that the user is in fact within that original environment (Macredie et al., 1996). The virtual environment is typically computer generated and controlled which provides various levels of interaction with the user. The use of work integrated learning and virtual reality as Bricken (1990) argues is capable of constructing a bridge which can assist the student to enhance their learning experience and ultimately achieving better results in their studies.

This report will discuss the WIL approach and VR as a modern technology and then examine the synthesis of both WIL and VR as WILVR. Furthermore, the characteristics as well as limitations of such a synthesis will be discussed. Finally, the report concludes with recommendations including identify potential areas of research and applications, and suggestions for stakeholders.

# 2. Work Integrated Learning

WIL has been in existence for quite some time in one form or another such as apprenticeships, traineeships and work experience programs. However, currently greater emphasize is been placed on the importance of student employability and higher education is increasingly asked to account for the success rates for graduates to gain employment (Eraut, 1994 cited in Orrell, 2004). These changing demands have created an expectation that the education sector will respond in innovative ways to meet both the learning needs and the career goals of all its students. Thus, the concept of WIL is maturing and gaining greater complexity and deeper dimensions.

2.1 Definition

The term Work Integrated Learning can be defined in several ways depending on the context, industry and scope of interest.

Patrick et al. (2008:9 cited in Smith & Raymond, 2010) defines WIL quite broadly as "an umbrella term for a range of approaches and strategies that integrate theory with practice of work within a purposefully designed curriculum".

Another broad definition of WIL provided by Martin & Hughes (2009) is that WIL represents a bridge connecting the student's academic present with their professional future; that it a chance to practice academic theory in authentic workplaces, and to develop real professional skills in order to prepare for their future career.

Meanwhile, Campbell (2011) emphasizes that the definition of WIL must be more inclusive involving, for example, practical experiences undertaken by learners as part of their internships and academic courses as well as simulations, real-life case investigations and classroom projects.

By contrast, Griffith University (2006 cited in McLennan & Keating, 2008) outlines more specific definition of WIL as intentional, meaningful and structured educational activities which assimilate theory with practical workplace application; and that the objectives of WIL must focused on teaching applied and transferable skills.

Finally, Moreland (2005, p. 4 cited in McIlveen et al., 2011) stresses the personal aspect of WIL and defines the term as an empowering approach which involves students learning about their prospective work and about themselves which would help them succeed in their future professional and personal lives. Gaining practical experience and even the learning process itself are seen as valuable tools for self management in future work situations.

Nonetheless, despite the variety of definitions, it is clear that WIL involves a combination of practical activities along with theoretical knowledge with the aim of preparing the learners for the real workplace and optimizing the chances of success in their careers.

2.2 Principles of WIL

The principles of Work Integrated Learning can be summarized as follows:
1) WIL is a three way relationship between the student, the academic institution and the workplace, requires all involved to perform specific tasks and accept certain responsibilities (Martin & Hughes, 2009).
2) WIL must be based on collaboration among stakeholders, negotiation in process and program design and transparency in objectives and methodology (Smith and Betts, 2000, p600-2 cited in Gibson et al., 2002).
3) Each WIL student should have an academic advisor and an industry supervisor for ongoing support and mentoring (Jancauskas et al., 1999).
4) WIL work should include interesting work activities and learning needs to be intentional and proactive to insure the high standards and desirable outcomes (Orrell, 2004).
5) WIL works best if there is equivalence in assessment. This means Awarding credit for WIL study units, equivalent to credit offered in pure academic units (Macquarie University, 2007 cited in McLennan & Keating, 2008).
6) Self-evaluation and other reflective activities are an important part of WIL programs to encourage a proactive outlook for the learners (Macquarie University, 2007 cited in McLennan & Keating, 2008).
7) WIL must be based on appropriate academic standards in teaching and assessment, and workplace placements must be in suitable contexts (Macquarie University, 2007 cited in McLennan & Keating, 2008).

2.3 Benefits of WIL to stakeholders

There are various stakeholders of Work Integrated Learning ranging from the social community to certain governmental departments. However the three main stakeholders are:

1) Industry

2) Students / Learners

3) Academic institutions

There are various benefits of Work Integrated Learning for the stakeholders:

2.3.1 Benefits to industry:

1) Enhancing academic-industry relations by supplying relevant industries with qualified and highly-skilled workers (Downey, 1979 cited in Rittichainuwat et al., n d).

2) WIL enlarges the scope of the professional educational curriculum and informs industry about new ideas and developments in education, as well as research advancements in the field (Martin & Hughes, 2009).

3) WIL provides employers an opportunity to scan for suitable prospective staff, ad to actively and directly participate in the training of those prospective employees (Martin & Hughes, 2009).

4) Developing a workforce capable of performing results oriented tasks with minimal need for extra job training (Marton, 1993 cited in Jackson, 2009).

2.3.2 Benefits to students / learners:

1) Providing students with academic and career benefits as they learn to utilize their academic knowledge effectively, adapt to workplace environments and solve real problems. There are also career benefits such as decision-making skills, improved self-confidence, career clarification, valuable work experience as well as developing personal attributes to increase employability and success in the workplace (Jacob, 2005 cited in Bohloko & Mahlomaholo, 2008).

2) WIL can, in contexts of situated learning, help indicate the degree of learner suitability or readiness for their chosen field of study (Hayes, 2006).

3) Learner experience is enriched through safe but highly challenging program activities (Harris et al, 2008).

4) Students can gain experience by having contact with professionals on the job, and at the same time discover within themselves areas of strengths which they can further develop and weaknesses which they can improve.

5) WIL programs can often give learners exposures to variety of work situations rather than a single workplace or particular organization, which helps develop well-adjusted learner views on the diversity of workplaces and help them appreciate the value of their future work (Crump, 2008).

2.3.3 Benefits to academic institutions:

1) Successful / profitable WIL projects, jointly owned between industry and academic institutions, can represent a useful income stream for the university, for example, the Virtual Art Gallery on a Second Life Island (Rossignol, n d).

2) WIL can bring valuable resources, including financial support in the form of grants for projects, to universities and other academic institutions which are essential for growth and capacity building (Rossignol, n d).

3) WIL helps in engaging teaching institutions closely in work realities and emerging practice norms (Harris et al, 2008).

2.4    Roles of stakeholders of WIL

2.4.1    Academic Institutions

The history of academic involvement in practical training is quite old, and Roodhouse (2007, cited in Gilbert, 2008) traces it to some of the oldest universities in Europe which had work-based learning programs. The role of academic institutions in WIL can be summarized as:

1) Academic institutions / universities represent the important venue where the teaching of the academic components of WIL programs takes place. However, universities often host the practical component of the training. For example, In Siam University's WIL in 1997, a training kitchen and restaurant were especially built for the project (Rittichainuwat et al., n d).

2) Universities must have flexibility in curriculum design and be responsive in tailoring WIL to suit individual project requirements (Brown, 2008).

3) Academic institutions are ideals for conducting formal assessment of WIL projects, as they have the resources and expertise for formal assessment (Jancauskas et al., 1999).

4) Universities can act as a significant bridge linking industry demands, student expectations and community needs. Davies (2007) explains that WIL must be seen to take place between "community partners" in order to be successful.

5) WIL projects can run as multi-disciplinary or even across many faculties within a university which means enhanced knowledge and understanding (Rossignol n d).

6) The accumulation of WIL outcomes in universities represents valuable learning tools and resources widely available for students and the wider community (Rossignol, n d).

7) Universities are ideal for internationalizing learning contexts and WIL can help multidirectional flow of knowledge and expertise with the result that both local and international students have an enriched learning experience (Davies, 2007).

8) Universities are not driven by the motive of maximizing profit as much as industry, so they can run not only WIL projects which benefit industry, but also society community focused projects, which would enhance students' skills required to build their "social capital" (Huq & Gilbert, 2009).

2.4.2 Industry

The role of industry and its various organizations is crucial for WIL and includes:

1) Industry usually provides the necessary funding for WIL projects.

2) It is the role of industry to specify the skills required in the market place (Reinhard, 2006).

3) Industry can lead project design, and provide valuable input into what constitutes efficient designs of WIL project facilities (Reinhard, 2006).

4) Providing necessary technical expertise, as well as state of the art technology is an important role of industry in WIL (Edwards, 2007).

5) Jointly (alongside academic staff) supervise the state of the WIL project as well as the progress of students (Edwards, 2007).

6) Expose students to real working environments and providing an opportunity to face real life challenges.

7) Teach students practical skills e.g. customer relationship management, professional ethics & successful communication skills (Martin & Hughes, 2009).

2.4.3 WIL issues for students / learners

Students are important stakeholders in WIL as they are the focus point around which the projects run. The issues of concern to students include:

1) WIL must have reasonable balance between theory and practice and be student focused rather than to serve the need of a specific company or organization. Getting broad skills is preferable to gaining company-specific qualifications (Reinhard, 2006).

2) Students can be concerned that their ability to work on the workplace may not be properly valued within assessment structures, which could affect students' focus and attention (Smith, 2010).

3) Students often indicated the final assessment report is too general and that there needs to be greater flexibility and detail in assessment reports (Smith et al., 2010).

4) Many students emphasize the value of creating shared learning spaces to support reflection and dialogue, so they can consider their experiences in the light of their peers' experiences (Smith et al., 2010).

5) The scope and team allocation of WIL projects need to be appropriately to the student's ability and needs. If projects are large, or if the team is not cohesive, then success is difficult to achieve (Langworthy, 2003).

6) Suitable work placement is often difficult to achieve. For example, in a case study by O'Reilly et al. (2010), less than 24% of students managed to find their own placement.

It is important to realize that in order to achieve success in WIL projects, all stakeholders must work collaboratively. They must share ideas, resources and expectations throughout the lifecycle of the project (Rittichainuwat et al., n d).

2.5 Challenges for WIL

Despite its widespread application and high popularity, WIL still faces some significant challenges:

1) WIL can present some issues for students including extra travel costs, expense involved in purchasing practical workplace equipment dress and limited opportunities to obtain placement in the learner's specific study area (Davies, 2007).

2) Industry professionals can face difficulties in managing students on WIL programs, giving them feedback and allocating space for students placed in the company, in addition to the pressure of supervising the students in their busy professional schedule (Davies, 2007).

3) WIL can be challenging for students in aspects such as cross-cultural communications, ethnic and gender differences, personal development issues, and values within a diverse workplace context.

4) Some WIL programs requires the students to prepare a resume, contact employers, prepare for and attend interviews, be short-listed and negotiate contracts regarding areas of work and salary and find accommodation often in a distant location. Such skills and competencies are sometimes not appropriately acknowledged as part of the assessment in WIL, and often there is little support, mentoring or guidance for those activities (Pitout, 2009).

5) WIL programs can suffer from insufficient monitoring of students in the employment situation where they are placed. Orie (2007, cited in Pitout, 2009) notes that, as a minimum, supervisory visit to the workplace must occur within six weeks of student placement and four times a year after that.

6) Not all subjects are suitable for the WIL model. For instance, Khan (2008) points to the difficulty of designing meaningful content, selecting appropriate delivery modes, and developing suitable assessment in highly theoretical subjects such as mathematics.

7) Skills and experience of academic staff is not always suitable for WIL programs and sometimes such skill lags behind industry standards (McLennan & Keating 2008).

8) WIL can be resource intensive for universities without apparent immediate tangible results (McLennan & Keating 2008).

Thus, it can be seen that, despites its long history and the various benefits of WIL, it still faces some significant challenges. Those challenges are of concern to all stakeholders involved including students, academic institutions and industry, and the issues are often interlinked. For example, the challenge of cross-cultural communications facing students, is in fact an important issue for universities too, while the issue of cost in implementing WIL programs affect academic institutions as much as industry.

## 3. Virtual Reality

Just like WIL, Virtual Reality (VR) is a technology that has been gaining increasingly popularity, research attention and application. However, it is a broad concept that is difficult to define. Since its inception, VR as cutting edge technology, has promised much in the various fields of modern life; some of which it has succeeded in delivering and some it has to yet fully accomplish.

3.1 Definition / Key ideas of VR

VR can have several close but separate definitions, depending on the application focus, research field or specialized interest of the context.

As an early researcher in the field, Steuer (1992) noted that the term VR was first used in 1989 and that it came from the IT industry and not academic institutions. He pointed out that VR was being defined as a "medium" composed of a hardware collection, including computers, headphones, display units and sensory gloves, but he criticized that definition as being too restrictive and over focus on technology itself instead of on the user. Thus, he suggested that a preferable definition would be that VR represents a special kind of experience assisted by an IT framework. As part of highlighting the importance of the user, Steuer (1992) emphasized the concepts of "presence" and "telepresence" as major aspects of that definition.

By contrast, other researchers, such as Brooks (1999) define VR more broadly as any experience "in which the user is effectively immersed in a responsive virtual world". Nonetheless most definitions of VR, including the two mentioned above, agree that it is an environment which simulates reality to variable degrees and includes various interactions with the user such as visual, audio and haptic (sense of touch).

## 3.2 Technologies (experiential / immersive)

The experiential aspect of VR environments has become more specific and the term "immersive" is been used to indicate a high degree of interaction between the user and the VR system. Immersion refers to having a simulated VR environment where the user feels immersed in the experience of interacting with that environment. In many cases there is much interest in fully immersive systems whereas in others, it is seen as unnecessary or even undesirable.

On the one hand, some like Wyld (2010) suggest that more immersive technologies are inevitable, that the popularity of the 3d virtual website Second Life is proof of that trend and that if companies are to remain competitive, then they must adopt some kind of interactive and immersive VR as a platform for marketing their products. Similarly, Linebarger et al. (2005 cited in Davis et al., 2009) found improved collaboration in virtual product design teams who operated in highly immersive VR environments.

In contrast, some warn against excessive immersion in VR and point to its detrimental effects. Some have even warned that with increasing use of VR, the foreseeable future would mean technology-enhanced people living their lives within virtual immersive worlds that they can select and adjust individually (Donovan, 2010).

There is of course also a middle position which advocates semi-immersion or a mixed reality environment. For example, Cheng & Fan (2010) point to the MMU (Manned Maneuvering Unit) project for training astronauts which offers a computer-simulated environment and provides a sense of partial immersion through head mounted displays and data-gloves, as a successful example of combining the physical and virtual environments. Another example is the virtual museum model based on a semi-immersive system due to cost issues as well as installation restrictions (Lepouras et al., 2003).

## 3.3 Pedagogical relevance

There are other applications of VR besides its well known uses in training such as for pilots in flight simulators. In fact there are many other fields utilizing or at least exploring the use of this technology as a valuable training platform:

One of the pioneering fields to exploit the use of VR for professional training is medicine. Zajtchuk & Satava (1997) point out that areas in the medical field which utilize VR technology as integral components of their training programs include casualty care and disaster planning, patient rehabilitation and psychotherapy, and they assert that medical simulators will become standard in medical training, testing and certification just as flight simulators are in the aviation industry today. They predict that medical students will be able to see the effects of surgical instruments on body organs and even be able to feel the pressure forces in their fingers through improved future immersive technology.

In addition, some VR projects are extending beyond their narrow scope or field of application and are helping in the wider educational context. For instance, Project ScienceSpace, which is a joint project between the NSF (National Science Foundation) and George Mason University, in the US, is providing multisensory environments which are interactive and highly immersive for physic and chemistry high school students (Sherman et l., 1997).

Thus, there is a wide range of VR application in training programs in a diverse range of fields.

## 3.4 Challenges for VR

Despite all the promising potential of VR, it still faces some notable challenges:

1) One significant drawback for VR systems is that they are often tied to a proprietary product which limits integration of components with other systems. Schiavenato (2009) gives an example of the Program for Nursing Curriculum Integration which necessitates expensive mannequins whose value is questionable according to the author.

2) High financial cost is another serious challenge for many VR systems, especially for those which use the latest and newest technology, and return on investment is not always immediate (Dickey, 1999).

3) Some VR applications do not provide adequate physical feedback, in the form of audio-visual or other sensory signals, to the user, and so offer less interactivity (Steuer, 1992).

4) Gutierrez-Maldonado et al. (2010) criticizes the portrayal of "perfect" looking people in some VR representations, and the emphasize on "body image" as this creates unrealistic or even impossible expectations in young people's minds about physical looks and body shapes.

5) Some studies question whether there is sufficient evidence of transfer of training skills from virtual reality to the real world, and that wide ranging studies are required to prove that transfer occurs (Cox et al., 2010).

6) Some researchers suggest that company managers should not assume that employees trained in a VR environment will behave in same way in the real world (Davis et al. 2009).

7) Design faults & errors (both hardware and software) as well as oversights by designers can compromise the experiential quality of a VR system and lead to a gap between the expectations/promise of VR and what is delivered (Hutchison, 2006).

8) There are situations that are very difficult to simulate with currently available technology, such as authentic modeling of human tissue that is responsive to surgical operations (Seymour & Røtnes 2006).

9) Despite VR environments including audio, visual and haptic senses, they still lack other senses such as smell and no taste and therefore cannot provide complete "vividness" of experience (Steuer, 1992).

## 4. Integrated VWIL

VWIL is a new and promising area of interesting to many, including industry professionals, educators and community and government organizations.

4.1 Strengths and positive characteristics

The positive features and characteristics of VWIL include:

**A – Authentic interactivity**: Due to the promise to offer a new, efficient and interactive alternative in the area of professional adult training, it is possible for authentic WIL experiences, according to some academic studies, to be recreated in a virtual environment (Davies & Shirley, 2007). For example, a trainee mechanic assembling some virtual car parts can hear squeaks if a mistake is made or the parts can push back, or even guide the trainee's hand to the correct assembly and offer audio feedback as a explanation (Bricken, 1990).

**B – Flexibility**: VWIL is flexible, self-paced and can be more easily modified to suit the needs of individual learner situations. It is seen as a successful platform for training and education. Many studies, including Bricken (1990) point out that VWIL can engage the learner's attention which is the first step in effective education. They refer to VWIL as personalized "programmable environments" which can accommodate user preferences. Moreover, Jones & McCann (2005) state that VWIL is useful for managers who may be under stress or travel much, as it combines the flexibility of e-learning but with the added advantage of real-life simulation.

**C – Constructivism**: WVIL systems are not limited to only viewing and the learner can do much more. According to Davies & Shirley (2007), the user can interact, control and manipulate the state of objects and use specialized tools to create new objects, modify existing ones, and to undergo simulated assessment tasks all within the VR environment.

**D – Wider availability**: As VWIL widens the availability for training, a greater number of learners can take part, and there are even international Virtual Work Integrated Learning programs in many fields today. Davies & Shirley (2007) point to the Queensland University of Technology's Law Faculty's international WIL program, The International Virtual Placement, which utilizes a virtual platform. The program is international and is centered on the concept collaborative global education where law students from different countries learn cooperatively and gain valuable international experience. The authors report that the increasing spread of globalization may necessitate such an approach, e.g. lawyers in one country many need to learn the system legal in another, so VWIL is seen as an ideal approach especially with the use of online chatting, video conferencing, discussion forums, blogs and wikis as some of the communication tools linking the users in different countries. Students in the program can apply for work placement with a broad range of potential international employers including government, law firms, community and industry (Davies & Shirley, 2007).

**D – Equal opportunity platform**: VWIL creates a level playing field where the learner's age, gender, physical features or social status are not used to discriminate for or against them in the virtual world (Bricken, 1991).

**E – Safe customizable, low cost environments**: In many industries, VWIL provides a controlled, simulated environment where the learner can accomplish tasks safely, and this is not just limited to pilot training on flight simulators. Van Wyk & de Villiers (2009) reports about how the South African mining industry is using VWIL and notes the following benefits:

1) Low cost alternative to creating a real-life full-scale model for worker training.

2) Improved safety awareness within the workforce.

3) Multilingual systems can communicate with trainees in the native language.

4) Can design and implement a diverse range of scenarios, even those seldom encountered in the actual work.

5) Can safely simulate hazardous situations to users without exposing them to any real danger.

6) The learner can review and evaluate their records, compare with peer results and asses their performance individually or as part of a team.

7) The VWIL instructors can easily identify problem areas for the learner and adjust the training program accordingly.

Therefore, VWIL has the benefits of being flexible, customizable and interactive program as well as offering an authentic experience to the learner or trainee.

4.2 Challenges for VWIL

Despite all the promising potential, there is still issues and challenges facing VWIL:

**A – Assessment validation**: It is important to investigate whether the experience gained in VR training transfers to comparable situations in other VWIL training tasks. More importantly, it is essential to verify whether that experience transfers to similar situations in the real workplace, and to determine the degree of transfer (Bricken, 1990).

**B – Cost Burden**: With VWIL systems, cost can be an issue, and especially the question of who should pay, industry, government or academic institutions, for initial system design and implementation. This becomes even more important as technology advances and systems become obsolete in relatively short time.

**C – Usability limitations**: It is often difficult to translate system requirements into applications which truly simulate the real world. For example, with some data gloves, gesture actions are limited by the number of hand and fingers movement combinations that can be defined as distinct, mutually exclusive, and unique in the system and that are comfortable and natural for the user (Bricken, 1991).

**D – Fear of technology**: Some people are technophobic when it comes to using a computer, let alone wearing a head mounted device, or an electronic glove. Studies suggest that this fear is often rooted in the fear of making a mistake, the fear of identity loss and the fear of confusion (Bricken, 1991).

**E – Pre-training education requirement**: It is important to insure that trainees on a VWIL program are prepared and ready, with theoretical training and expectations, before placing them in VR environments. For example, Gallagher et al. (2005) notes that it would be pointless to place a junior surgeon in a fully robotic environment and stresses that training strategies in any VWIL field must insure that learners have a good idea of what to do, what not to do, where and when to do it, and why to do it.

**F – Virtual community issues**: In professional virtual learning communities (VLC), participants often have issues with time management, communication difficulties (e.g. communicating in a foreign language or slow typing speed) which highlights the asynchronous nature of participation in such communities. Moreover, access to ICT facilities, individual motivation, balancing work and life commitments, and deciding on private/public boundaries are also important factors which determine the success of such communities (Dickey, 1999).

**G – Trainer readiness**: Some industry instructors and educational institutions are slow to embrace the concept of VWIL because of a variety of reasons including cost and complexity especially with budget limitations in public education. Furthermore, some teachers/instructors do not have sufficient suitable IT skills to be involved in VWIL programs, and some may also be technophobic (Skylar, 2007).

Thus, VWIL faces some notable challenges including the issue of limited delivery modes, cost of design and installation and the fear of technology which many people face.

## 5. Discussion

As already outlined, VWIL offers advantages but also faces some challenges. It is a promising technology in many fields, especially in the medical, aviation and aerospace and military. However, it is not limited to those fields, and even law students in some universities are taking part in VWIL programs and reporting positive results. Yet, at the same, VWIL poses challenges for its stakeholders. The question of cost is a legitimate question, and this relates to availability too. For instance, a VWIL program which costs a lot of money to design and implement, may be out of reach for middle or lower income learners because the fees may be too high for them. Moreover, such high costs may make government bodies, community organizations or even the industry, reluctant to participate. Another related issue which needs to be addressed is that of value. It is difficult to quantify the value delivered by a VWIL program in simple numbers. The real value of such programs cannot be measured in terms of profitability or number of learners who can pass the exams. Furthermore, it is likely that the true value delivered by VWIL programs is long term rather than immediate and that it is of direct benefit to the individual learner, but also of benefit to the wider community. At the same time, the assessment methods employed in VWIL need to be detailed in outlining areas of strengths, achievements and specific experience and skills gained in such programs. There is no standardization of assessment in VWIL, and this makes it difficult to align qualifications to some common assessment scale or system. Another area of interest, not explored in current research, is the potentially positive impact of the increasing application of VWIL on the environment. VWIL is reusable and much more resource and energy efficient than creating real physical training environments in many situations. This makes VWIL a more sustainable and environment friendly approach to workplace training. Such issues are of great concern in the world today, especially to younger generations who are also more familiar with the use of IT in their daily lives.

Thus, there are valid issues of concern regarding the costs involved, the value delivered by VWIL programs and the current assessment approached to VWIL, and these issues must be clearly analyzed and addressed to make VWIL a more acceptable, popular and successful approach in the future. At the same time, there are potentially promising features of VWIL, such as to promote its environmentally friendly characteristics and its appeal for current generations.

## 6. Recommendations

1) As a modern integrative technological approach, VWIL can be used in many fields, and the degree of its use can be varied, from complete fully immersive programs to assistive separate modules, depending on the situation and the training needs.

2) It is important for VWIL program managers to specify appropriate skill transfer assessment criteria and to follow up in the real workplace.

3) Learner training and preparation prior to placement in VWIL is important to insure a higher chance of success in the program.

4) Involving the learner (end user) from the start of VWIL program design can help in making these programs more user friendly and hence more experiential and immersive.

5) Safety of learners and instructors must be paramount in any VWIL program, so it is important that sensory communication (e.g. 3d vision) in such programs does not cause any harmful effects to the learner.

6) VWIL should be promoted as a distinct study field & specialized instructors can be trained in VWIL as a subject in its own right.

7) Greater community engagement can increase awareness and reduce misinformation about VWIL. For example, field trips for school students can present VWIL in a realistic way and help reduce technophobia as well as help students understand what VWIL is and what it is not.

8) Further research is recommended in some areas. For example, VWIL is likely suitable for many training situations where hands on experience is valuable, e.g. TAFE traineeships, but is probably not quite suitable for other professions where human to human interaction is the most important factor such as in sales, general office work , acting or team sport.

9) Future research can also be carried out to investigate the suitability of WIL for people with learning difficulties because it is rich in hands on tasks and may make it easier for those learners to remain focused than purely theoretical classes.

## 7. Conclusion

In conclusion, this report has outlined the concepts of WIL, VR and VWIL, and presented some of their main aspects. VWIL as a new technological approach is being increasingly used in many fields today, ranging from aerospace to medicine. There are several benefits offered by VWIL including flexibility, customizability and the provision of an authentic simulated environment in which the learner can gain valuable practical workplace experience. Nevertheless, VWIL also faces some challenging issues including usability limitations and cost, which must be properly analyzed and addressed.

**Waleed Al Shehri** received his bachelor degree in computer science from King Abdulaziz University, Jeddah, Saudi Arabia (2005), MSc degree in information technology form Macquarie university, Sydney, Australia (2011). His current research interests in databases and software engineering . Currently working in the Department of IT in Royal Saudi Air Force ( RSAF ).